\documentclass{emulateapj}
\usepackage{amsmath}
\usepackage{natbib}
\usepackage{graphicx}
\usepackage{epstopdf}
\usepackage{txfonts}
\usepackage{gensymb}
\usepackage{color}
\usepackage{CJKutf8}

\newcommand{\rp}{r_{\rm p}}

\newcommand{\eqnref}[1]{Equation \ref{#1}}
\newcommand{\secref}[1]{Section \ref{#1}}
\newcommand{\figref}[1]{Figure \ref{#1}}
\newcommand{\tabref}[1]{Table \ref{#1}}

\shorttitle{Inferring Planet Mass from Spiral Structures}
\shortauthors{Fung \& Dong}

\begin{document}
\begin{CJK*}{UTF8}{bsmi}
\title{Inferring Planet Mass from Spiral Structures in Protoplanetary Disks}
\author{Jeffrey Fung (馮澤之)\altaffilmark{1} \& Ruobing Dong (董若冰)\altaffilmark{2}}
\affil{Department of Astronomy, University of California at Berkeley, Campbell Hall, Berkeley, CA 94720-3411}
\altaffiltext{1}{NSERC Fellow}
\altaffiltext{2}{Hubble Fellow}

\email{jeffrey.fung@berkeley.edu}

\begin{abstract}
Recent observations of protoplanetary disk have reported spiral structures that are potential signatures of embedded planets, and modeling efforts have shown that a single planet can excite multiple spiral arms, in contrast to conventional disk-planet interaction theory. Using two and three-dimensional hydrodynamics simulations to perform a systematic parameter survey, we confirm the existence of multiple spiral arms in disks with a single planet, and discover a scaling relation between the azimuthal separation of the primary and secondary arm, $\phi_{\rm sep}$, and the planet-to-star mass ratio $q$: $\phi_{\rm sep} = 102\degree (q/0.001)^{0.2}$ for companions between Neptune mass and 16 Jupiter masses around a 1 solar mass star, and $\phi_{\rm sep} = 180\degree$ for brown dwarf mass companions. This relation is independent of the disk's temperature, and can be used to infer a planet's mass to within an accuracy of about 30\% given only the morphology of a face-on disk. Combining hydrodynamics and Monte-Carlo radiative transfer calculations, we verify that our numerical measurements of $\phi_{\rm sep}$ are accurate representations of what would be measured in near-infrared scattered light images, such as those expected to be taken by Gemini/GPI, VLT/SPHERE, or Subaru/SCExAO in the future. Finally, we are able to infer, using our scaling relation, that the planet responsible for the spiral structure in SAO 206462 has a mass of about 6 Jupiter masses.

\end{abstract}


\section{Introduction}
\label{sec:intro}
Young, forming planets embedded in protoplanetary disks are expected to generate global spiral structures that are potentially observable. Recent observations show a variety of spiral structures in the scattered light images of protoplanetary disks, such as those in HD 100453 \citep{Wagner2015}, MWC 758 \citep{Grady2013, Benisty2015}, SAO 206462 \citep{Muto2012, Garufi2013}, HD 142527 \citep{Casassus2012, Rameau2012, Canovas2013, Avenhaus2014a, Christiaens2014}, and HD 100546 \citep{Boccaletti2013, Avenhaus2014b, Currie2014}. In all cases, the observed disk contains more than a single spiral arm, and while this may point to the existence of multiple planets, it has been shown that a massive planet alone can excite multi-armed spiral structures; e.g., \citet{Dong2015b} found that the two-armed spirals in both MWC 758 and SAO 206462 are consistent with the influence of a single massive planet.

In the foundational theory of disk-planet interaction \citep{GT1979, A1993}, planets excite density waves in disks over a range of $m^{\rm th}$ order azimuthal modes, and these waves interfere to create an one-armed spiral structure \citep{Ogilvie2002}. This picture is based on the assumption that the planet's perturbation on the disk is sufficiently small, so that the perturbation equations can be linearized. Typically this refers to planets whose ``thermal mass'', defined as $q_{\rm th}\equiv q/H_0^3$, where $q$ is the planet-to-star mass ratio, and $H_0$ the disk aspect ratio at the planet's location, is much less than unity. If we consider a more massive planet with $q_{\rm th}\sim 1$, then it can also excite ``ultraharmonic'' density waves, generating a secondary, or even tertiary, set of waves that can interfere with the primary set to create multi-armed structures.

Ultraharmonic wave excitation can be understood as the result of high order coupling between perturbations of different $m^{\rm th}$ azimuthal modes, and is a natural way to extend the conventional perturbation theory into the weakly nonlinear regime. Originally developed under the context of galactic dynamics, \citet{Artymowicz1992} showed that the quadratic perturbation terms, or the self-coupling terms, dropped in the linear analysis by \citet{GT1979} are in fact the driving terms for a second order perturbation that excites density waves at the 2:1 resonance with the primary arm. This can be generalized to even higher order perturbations, such as a third order perturbation at the 3:1 resonance, driven by the coupling between the first and second order terms. Evidence for these features may have already been found in previous work, such as \citet{Zhu2015}, who reported and characterized a secondary and tertiary arm in their simulations of disk-planet interaction; and \citet{JBP2015}, who also reported the existence of a secondary arm when $q_{\rm th}\gtrsim 1$. 

These spiral features are potential diagnostics for inferring disk and planet properties; for instance, \citet{Zhu2015} proposed the use of pitch angle, spiral separation, and spiral contrast to infer planet mass. Despite that, the properties of multi-armed structures remain poorly understood, due to the complexity of its nonlinear nature. In this paper, we investigate how the shape and location of the secondary arm vary with planet mass and disk temperature, with an emphasis on its applications to protoplanetary disk observations. Throughout this study, we focus on the part of the disk interior to the planet, because the outer disk is usually much fainter than the inner disk and less likely to be observed. We defer the thorough investigation of ultraharmonic resonances in disk-planet interaction to a future paper.

\section{Numerical Method}
\label{sec:numerics}
As the basis for our analysis, we perform global 2D disk-planet interaction simulations using the Graphics Processing Unit-accelerated hydrodynamics code \texttt{PEnGUIn} \citep{mythesis}. \texttt{PEnGUIn} solves the Navier-Stokes equations for a compressible fluid in the Lagrangian frame:
\begin{align}
\label{eqn:cont_eqn}
\frac{D\Sigma}{Dt} &= -\Sigma\left(\nabla\cdot\mathbf{v}\right) \,,\\
\label{eqn:moment_eqn}
\frac{D\mathbf{v}}{Dt} &= -\frac{1}{\Sigma}\nabla P + \frac{1}{\Sigma}\nabla\cdot\mathbb{T}  - \nabla \Phi \,,
\end{align}
where $\Sigma$ is the disk surface density, $\mathbf{v}$ the velocity field, $P$ the vertically averaged gas pressure, $\mathbb{T}$ the Newtonian stress tensor, and $\Phi$ the combined gravitational potential of the star and the planet. We adopt a locally isothermal equation of state, such that $P= c^2 \Sigma$, where $c$ is the sound speed of the gas. $\mathbb{T}$ is proportional to the kinematic viscosity $\nu$, which we parameterize using the $\alpha$-prescription by \citet{alpha}, such that $\nu=\alpha c^2/\Omega$, where $\Omega$ is the orbital frequency of the disk. In all our simulations, we choose $\alpha=0.001$, and verify by experimenting with a 10 times higher $\alpha$ that the specific choice of $\alpha$ has negligible effects on the spiral structure of the disks. 

Technically, the simulation is run in the corotating frame of the planet, but the Coriolis force is absorbed into the conservative form of \eqnref{eqn:moment_eqn} as suggested by \citet{Kley98}. We solve the equations in the rest frame of the star, such that the star is always stationary at $r=0$. This allows $\Phi$ to be written as:
\begin{equation}
\label{eqn:potential}
\Phi = -GM\left[\frac{1-q}{r}-\frac{q}{\sqrt{r^2 + r_{\rm s}^2 + \rp^2 - 2r\rp\cos{\phi'}}} - \frac{qr\cos{\phi'}}{\rp^2}\right] \,,
\end{equation}
where $G$ is the gravitational constant, $M$ the total mass of the star and the planet, $\rp$ the semi-major axis of the planet's orbit, $r_{\rm s}$ the softening length of the planet's potential, and $\phi' = \phi-\phi_{\rm p}$ denotes the azimuthal separation from the planet. The third term in the in bracket of \eqnref{eqn:potential} is the indirect potential introduced due to the acceleration of the frame. $r_{\rm s}$ is chosen to be $0.5H_0\rp $, appropriate for approximating the vertically averaged gravitational force \citep{Muller12}. We fix the planet to a circular orbit (i.e., planetary migration is ignored), so it is always positioned at $(\rp,\phi_{\rm p})$. We set $GM=1$, and $\rp=1$, such that the planet's orbital frequency $\Omega_{\rm p}$ is $1$, and orbital period $P_{\rm p}$ is $2\pi$. For convenience, we also denote the Keplerian orbital speed $v_{\rm k}$ and frequency $\Omega_{\rm k}$.

\subsection{Disk Model}
\label{sec:disk}
Our disk model uses a power law profile for the initial surface density:
\begin{equation}
\label{eqn:sigma}
\Sigma_0 = \Sigma_{\rm p} \left(\frac{r}{\rp}\right)^{-1} \,,
\end{equation}
where $\Sigma_{\rm p}$ is set to 1. \footnote{Since we do not consider the self-gravity of the disk, this normalization has no impact on our results.} Our choice of a fixed sound speed profile is:
\begin{equation}
\label{eqn:cs}
c = c_0 \left(\frac{r}{\rp}\right)^{-1/4} \,,
\end{equation}
This sound speed profile creates a disk with a flaring shape, since the disk aspect ratio $H=c/v_{\rm k}$ is proportional to $r^{1/4}$. The normalization $c_0$ is determined in terms of the aspect ratio at the planet's location $H_0 = c_0/v_{\rm k}(\rp)$. Finally, we complete our initial conditions with the velocity field $\mathbf{v}$. We set the initial radial velocity to zero everywhere, and the angular velocity to:
\begin{equation}
\label{eqn:omega}
\Omega = \sqrt{\Omega_{\rm k}^2 + \frac{1}{r}\frac{\partial P}{\partial r}} \,,
\end{equation}

To obtain a thorough picture of the planet-excited spiral arms, we survey over two parameters: $q$ and $H_0$. We vary $q$ from $6.25\times10^{-5}$, just under Neptune mass, to $6.4\times10^{-2}$, around brown dwarf mass; and $H_0$ from $0.04$ to $0.16$. For reference, we note that Jupiter has a $q$ of $10^{-3}$, and a change in $H_0$ can be interpreted as a change in the planet's location --- $H_0=0.1$, for instance, is equivalent to the planet being placed at about 25 AU away from a 0.5 solar mass ($M_\odot$) star in a passively heated disk (\citet{Chiang97}; also see Table 4 of \citet{Andrews2011} for disk scale heights in observed systems).

\subsection{Setup}
\label{sec:setup}

Our computational grid spans $r=\{0.125\rp, 2\rp\}$, and covers the full $2\pi$ range in azimuth. The azimuthal boundary conditions are periodic, and the radial boundaries contain fixed values as given by Equations \ref{eqn:sigma} to \ref{eqn:omega}. For cases when $H_0\geq0.1$, we use a resolution of $400(r)\times900(\phi)$, with the cells distributed logarithmically in the radial direction, and uniformly in azimuth. This gives a cell size of $\sim 0.007r$ in both directions. For models with $H_0<0.1$, we increase the resolution to $600(r)\times1400(\phi)$, so that even for the smallest $H_0$ used, 0.04, we still retain a resolution of more than 8 cells per scale height near the planet. 

The spiral structure is typically fully established after about one sound-crossing time, which, for our models, is about 1 to 4 $P_{\rm p}$. We therefore run all our simulations for a minimum of $t_{\rm end} = 10 P_{\rm p}$, and we verify that in most cases, the spiral structure is steady after 5 $P_{\rm p}$ or less. Since the more massive planets open a gap over a viscous timescale much longer than $P_{\rm p}$, we additionally run our $q=0.001$ and $H_0=0.1$ model to 300 $P_{\rm p}$ until the gap has fully opened, and confirm this does not alter the morphology of the spirals significantly. \footnote{At 300 $P_{\rm p}$, we measure $\phi_{\rm sep}$, the azimuthal separation of the primary and secondary arms (see \secref{sec:results}), to be $104\degree$, which is similar to $99\degree$ at 10 $P_{\rm p}$ reported in \tabref{tab} or $102\degree$ in \eqnref{eqn:bestfit}.}

For models with $q>0.001$, where the strong influence from the planet generates fluctuations that require a longer time to damp out. In these cases, we run the simulations for 20 $P_{\rm p}$ when $q=0.004$, and extend it to 30 $P_{\rm p}$ when $q\geq0.016$. These very massive companions are also able to carve out a deep gap within the timescale of our simulations, so the reported spiral morphologies of these models do include the effects of gap opening.

We additionally perform a 3D simulation to compare with our 2D results. It uses a similar setup as our 2D simulations, and the differences are detailed in \secref{sec:3D}.

\section{Results}
\label{sec:results}

\begin{deluxetable}{lclccl}
\tabletypesize{\footnotesize}
\tablecolumns{14}
\tablewidth{0pc}
\tablecaption{Simulated Models}
\setlength{\tabcolsep}{0.07in}
\tablehead{
\colhead{$H_0$} & \colhead{$q$} & \colhead{$\phi_{\rm sep}$~(degree)} & 
\colhead{$q_{\rm th}$} & \colhead{$t_{\rm end}~(P_{\rm p})$} & \colhead{\{$r_1$,~$r_2$\}~($a$)}
}
\startdata
0.16 & $6.25\times 10^{-5}$ & -~-~\tablenotemark{a} & 1/64 & 10 & -~- \\
0.16 & $2.5\times 10^{-4}$ & -~-~\tablenotemark{a}  & 1/16 & 10 & -~- \\
0.16 & $1.0\times 10^{-3}$ & $110~(+6)~(-5)$ & 1/4  & 10 & \{0.4, 0.6\} \\
0.16 & $4.0\times 10^{-3}$ & $125~(+3)~(-4)$  & 1   & 20 & \{0.4, 0.6\} \\
0.16 & $1.6\times 10^{-2}$ & $179~(+22)~(-19)$ & 4  & 30 & \{0.4, 0.55\} \\
0.16 & $6.4\times 10^{-2}$ & $178~(+6)~(-7)$  & 16  & 30 & \{0.4, 0.55\} \\
\tableline\\[-5pt]
0.1 & $6.25\times 10^{-5}$ & -~-~\tablenotemark{a}  & 1/16 & 10 & -~- \\
0.1 & $2.5\times 10^{-4}$ & $81~(+8)~(-6)$    & 1/4 & 10 & \{0.4, 0.7\} \\
0.1 & $1.0\times 10^{-3}$ & $99~(+10)~(-9)$  & 1    & 10 & \{0.4, 0.7\} \\
0.1 & $4.0\times 10^{-3}$ & $125~(+21)~(-18)$ & 4   & 20 & \{0.4, 0.65\} \\
0.1 & $1.6\times 10^{-2}$ & $172~(+6)~(-11)$ & 16   & 30 & \{0.4, 0.55\} \\
0.1 & $6.4\times 10^{-2}$ & $195~(+4)~(-2)$  & 64   & 30 & \{0.4, 0.5\} \\
\tableline\\[-5pt]
0.063 & $6.25\times 10^{-5}$ & $56~(+6)~(-7)$   & 1/4 & 10 & \{0.4, 0.7\} \\
0.063 & $2.5\times 10^{-4}$ & $75~(+8)~(-15)$   & 1   & 10 & \{0.4, 0.7\} \\
0.063 & $1.0\times 10^{-3}$ & $102~(+11)~(-15)$ & 4   & 10 & \{0.4, 0.7\} \\
0.063 & $4.0\times 10^{-3}$ & $143~(+19)~(-57)$ & 16  & 20 & \{0.4, 0.6\} \\
0.063 & $1.6\times 10^{-2}$ & $183~(+20)~(-22)$ & 64  & 30 & \{0.4, 0.5\} \\
\tableline\\[-5pt]
0.04 & $6.25\times 10^{-5}$ & $60~(+23)~(-12)$ & 1   & 10 & \{0.4, 0.7\} \\
0.04 & $1.0\times 10^{-3}$ & $100~(+19)~(-22)$ & 16  & 10 & \{0.4, 0.7\} \\
0.04 & $1.6\times 10^{-2}$ & $173~(+13)~(-30)$ & 256 & 30 & \{0.4, 0.6\}
\enddata \label{tab}
\tablenotetext{a}{The secondary arms in these models are too weak to be precisely located.}
\end{deluxetable}

\begin{figure*}[]
\centering
\includegraphics[width=1.99\columnwidth]{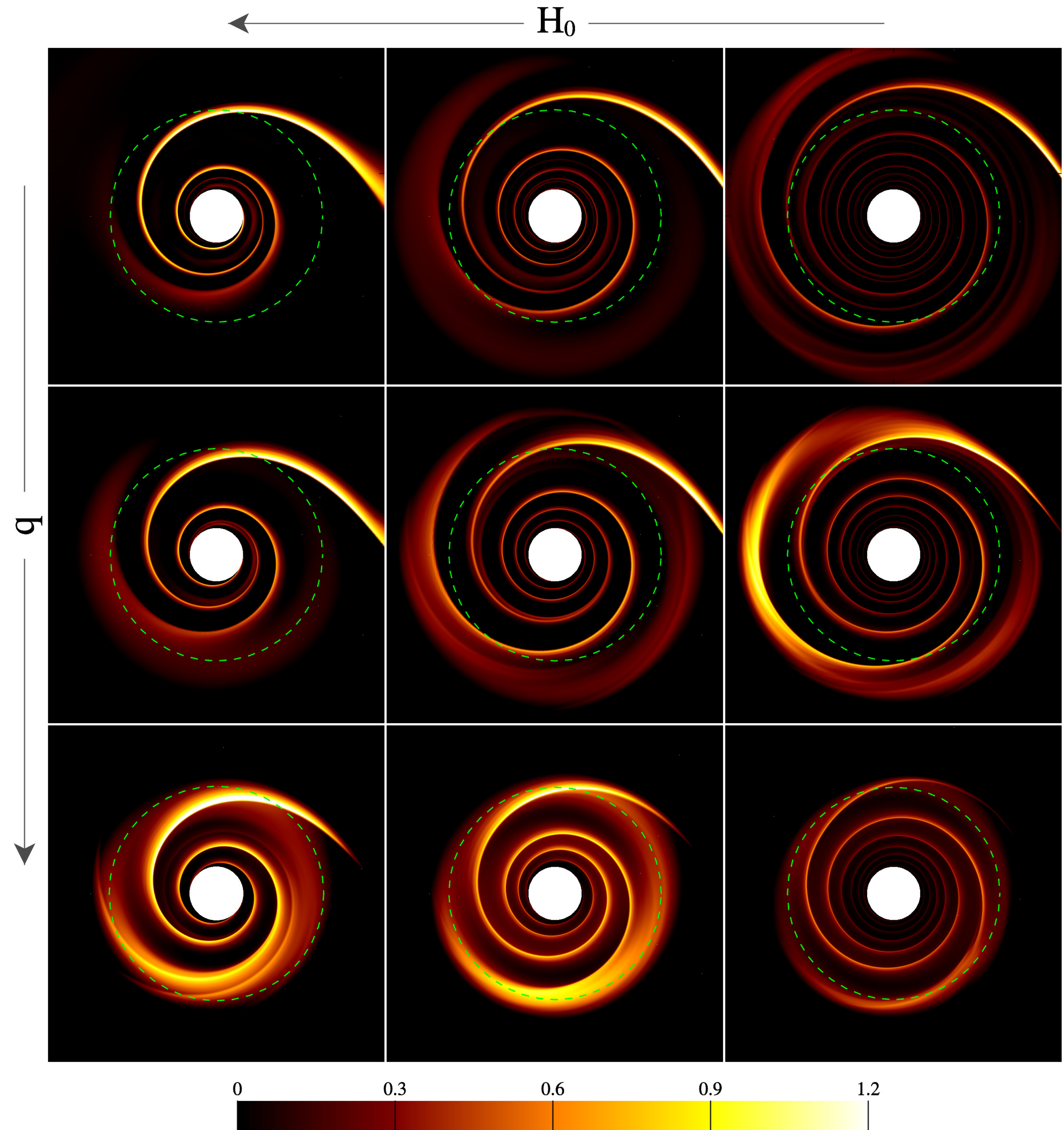}
\caption{Snapshots at the end point of 9 selected simulations. The color shows the normalized fractional variations in surface density $\Delta$, as defined by \eqnref{eqn:delta}. $H_0$ increases from right to left, where the left column contains snapshots of models with $H_0=0.16$, middle column has $H_0=0.1$, and right column has $H_0=0.063$. $q$ increases from top to bottom, where the top row has $q=0.001$, middle row has $q=0.004$, and bottom row has $q=0.016$. The green dashed circles have a radius of $0.5~\rp$, and are provided as a length scale. All snapshots has the same orientation where the planet is located to the right of the picture, outside of the frame; see \figref{fig:scat_light} for a zoomed-out version that includes the planet.}
\label{fig:montage}
\end{figure*}

\begin{figure*}[]
\centering
\includegraphics[width=1.99\columnwidth]{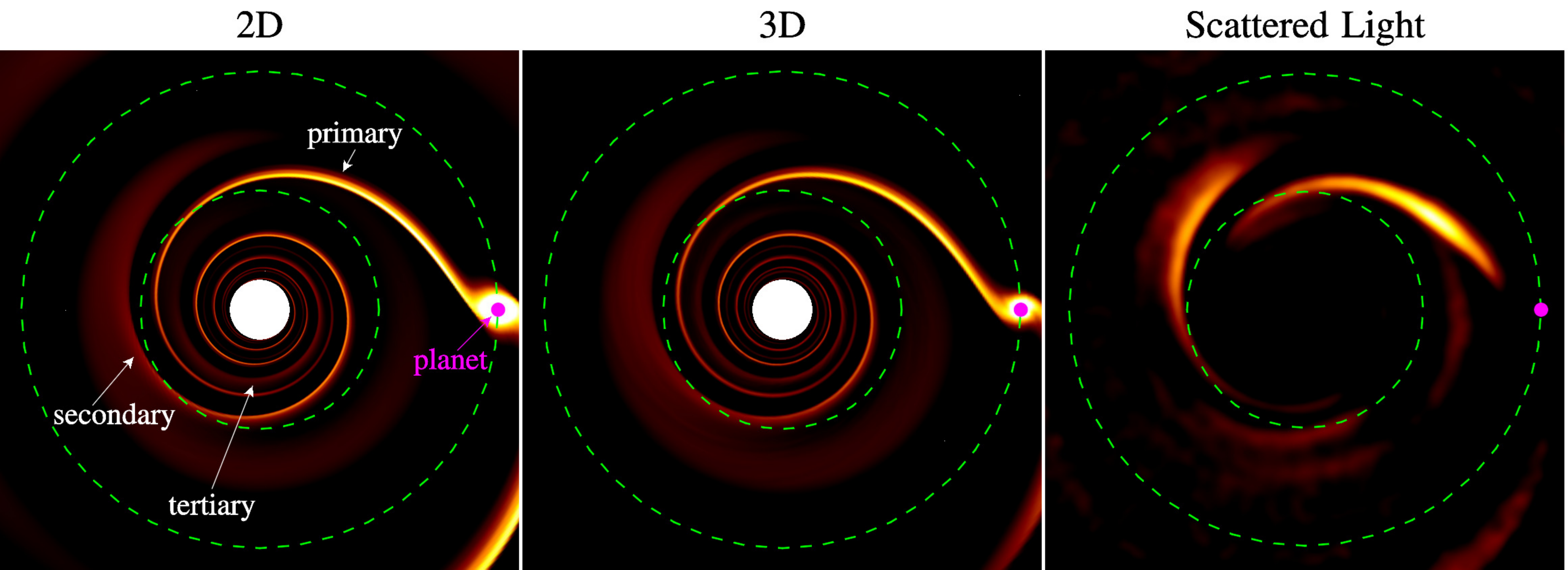}
\caption{The left panel is the 2D $\Delta$ (\eqnref{eqn:delta}) profile of the $q=0.001$, $H_0=0.1$ model, same as the top middle panel of \figref{fig:montage}. The middle panel shows the $\Delta$ map for the same model, but from a 3D simulation, and the right panel shows a face-on scattered light image corresponding to the 3D simulation. See \secref{sec:3D} for the details of this 3D model. Similar to the way surface density is rescaled into $\Delta$, we also rescale the scattered light image by a factor of $r^2$ to level the brightness profile. The magenta dot marks the location of the planet, and the concentric green dashed circles have radii of $0.5$ and $1~\rp$.}
\label{fig:scat_light}
\end{figure*}

We perform a total of 20 simulations with different sets of $\{q,~H_0\}$ parameters. The results are recorded in \tabref{tab}, and \figref{fig:montage} shows a montage of snapshots illustrating the surface density profile of 9 different models. In these maps, we plot the normalized fractional variation in the surface density, $\Delta$, in reference to the initial profile (\eqnref{eqn:sigma}):
\begin{equation}
\Delta = \frac{1}{A}\frac{\Sigma-\Sigma_0}{\Sigma_0} \,,
\label{eqn:delta}
\end{equation}
where the normalization $A$ is:
\begin{equation}
A = \left(1+3~q_{\rm th}\right)^{\frac{1}{3}}-1 \,.
\label{eqn:amplitude}
\end{equation}
This choice of $A$ is motivated by the notion that in the linear limit of $q_{\rm th}\ll1$, the amplitude of the density perturbation is well described by $A\approx q_{\rm th}$, while in the nonlinear, $q_{\rm th}>1$, regime, we speculate that the amplitude scales as $r_{\rm H}/(H_0\rp)\propto q_{\rm th}^{1/3}$, the ratio between the planet's Hill radius $r_{\rm H}$ and the disk scale height at the planet's orbit. Our prescription for $A$ connects these two regimes to provide a smooth scaling across our sample. As can be seen in \figref{fig:montage}, this amplitude scaling matches our results well.

Ultraharmonic theory predicts that the second order perturbation is excited at the 2:1 resonance with respect to the first order perturbation, or equivalently, the planet, since the primary arm corotates with the planet. Consequently, if the secondary arm is a product of the second order perturbation, we expect to find it being launched at $r/\rp\sim 0.63$. One can even predict a tertiary arm being launched at the 3:1 resonance, or $r/\rp=0.48$. On the left panel of \figref{fig:scat_light}, as well as the snapshots in \figref{fig:montage}, we do find the secondary and tertiary arms launching at locations similar to these predicted values. This is evidence that these arms are likely related to ultraharmonic resonances. In the weakly nonlinear regime ($q_{\rm th}\lesssim 1$) where ultraharmonic theory applies, the amplitude of the $n^{\rm th}$ order perturbation decreases with increasing $n$ following $q_{\rm th}^n$. Accordingly, we find the tertiary and higher order arms are typically too weak to be seen. In the fully nonlinear case ($q_{\rm th}>1$), we find that the tertiary arm also cannot be easily identified, since it appears to collide and merge with the primary. Hence, in this paper, we will focus only on the morphology of the primary and secondary arm.

Two clear trends can be identified from \figref{fig:montage}: 1) increasing pitch angle as the disk aspect ratio increases (right to left in \figref{fig:montage}), and 2) increasing the azimuthal separation between the two arm, which we denote $\phi_{\rm sep}$. as planet mass increases (top to bottom in \figref{fig:montage}). The trend in pitch angle is predicted by linear perturbation theory, where the pitch angle at large radial separations from the planet can be approximated as $\tan^{-1}(c/[r|\Omega_{\rm k}-\Omega_{\rm p}|])$ \citep[e.g.][]{Rafikov2002}. We note that, on top of this trend, there is an additional, weaker dependence on the planet mass, which is the consequence of the propagation of nonlinear ultrasonic wave \citep{Rafikov2002}. Therefore, it is possible to evaluate planet mass using measurements of the pitch angle, but there are two challenges with this method: 1) a priori knowledge of the $H$ profile of the disk is needed to model the pitch angle; and 2) the pitch angle is typically a strong function of its distance to the planet, and so a priori knowledge of the location of the planet is also necessary. For these reasons, we consider the trend in $\phi_{\rm sep}$ a more promising method for measuring planet mass.

\subsection{$\phi_{\rm sep}$ dependence on planet mass}
\label{sec:sep}

\begin{figure*}[]
\includegraphics[width=1.99\columnwidth]{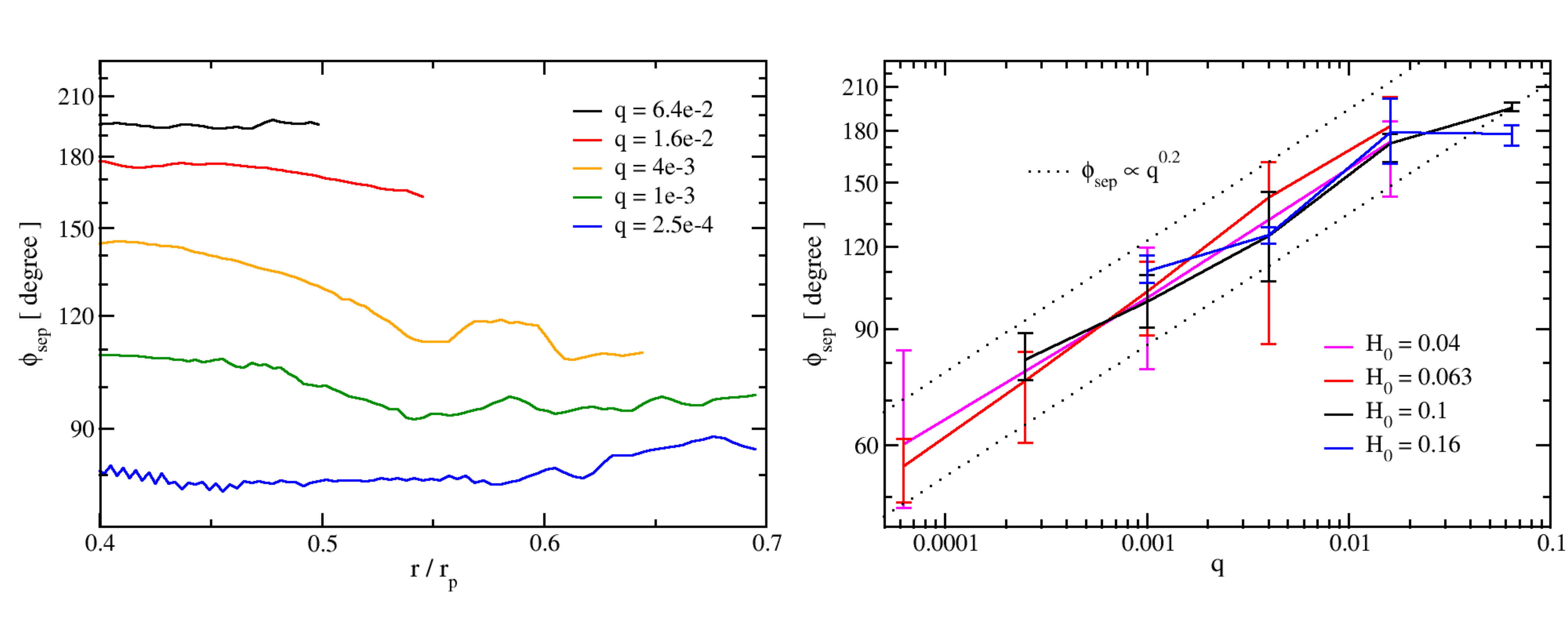}
\caption{The left panel shows $\phi_{\rm sep}$ as a function of the distance from the star for models with $H_0=0.1$. For more massive planets, the secondary arm is not seen at larger radii, due to a deep disk gap opened by the planet. The right panel is the average azimuthal separation $\phi_{\rm sep}$ between the primary and secondary arm, as a function of the planet-to-star mass ratio $q$. The dotted lines are constant $\phi_{\rm sep}\propto q^{0.2}$ slopes.}
\label{fig:sep}
\end{figure*}

For a given annulus, we locate the azimuthal positions of the primary and secondary arm by finding the local maxima in $\Sigma$ associated with the two arms, which then allows us to construct $\phi_{\rm sep}$ as a function of $r$. On the left panel of \figref{fig:sep}, we give examples of $\phi_{\rm sep}(r)$ using our set of models with $H_0=0.1$. Typically, $\phi_{\rm sep}$ fluctuates by about $\pm10\%$ over the range of $r/\rp=\{0.4,~0.7\}$, which is sufficiently small for us to clearly identify an increasing trend in $\phi_{\rm sep}$ as $q$ increases. In \tabref{tab}, we report $\phi_{\rm sep}$ as the average over the radial range $\{r_1,~r_2\}$, where $r_1$ is fixed at $r/\rp = 0.4$, while $r_2$ is the largest $r$ at which the secondary is still clearly defined, up to $r/\rp = 0.7$. We also report the maximum and minimum values of $\phi_{\rm sep}$ within this range as error margins.

The right panel of \figref{fig:sep} plots all of our $\phi_{\rm sep}$ measurements as a function of $q$. We find a well-defined correlation between the two that is independent of $H_0$. By inspecting Figure 2 and 3 of \citet{Zhu2015}, we find our $\phi_{\rm sep}$ measurements to be consistent with both their isothermal and adiabatic simulations, which lends confidence to the accuracy of our results, and also implies that $\phi_{\rm sep}$ is not only insensitive to $H_0$, but may also be insensitive to the equation of state of the gas.

To further test this apparent insensitivity to disk thermodynamics, we perform two additional $\{q,~H_0\} = \{0.001,~0.1\}$ simulations, one with a globally isothermal equation of state, $c = c_0$, which results in a more flared disk, and another one with $c = c_0 \sqrt{\rp/r}$, which results in no flaring. We find $\phi_{\rm sep}=103\degree$ and $96\degree$ respectively. This suggests the dependence on the radial disk temperature profile is also negligibly weak compared to that on $q$.

Overall, $\phi_{\rm sep}$ is proportional to $q^{0.2}$, up to $q=0.016$, beyond which $\phi_{\rm sep}$ remains approximately constant at $180\degree$. A fit using $\chi^2$ minimization gives:
\begin{align}
\nonumber
q \leq \, 1.6 \times 10^{-2} \,:& \\
\label{eqn:bestfit}
\phi_{\rm sep} =& \, 102\degree \left(\frac{q}{0.001}\right)^{0.2} \,,
\end{align}
which typically matches our results to within $5\%$. We note that $\phi_{\rm sep}$ is the separation measured face-on. In general, observed systems will have some inclination that must be taken into account before our scaling relation can be applied. Using the uncertainty in our fitted parameters, and assuming $\phi_{\rm sep}$ is only a function of $q$, we also derive $\Delta q$, the uncertainty in $q$. Given $\Delta \phi_{\rm sep}$, the measurement error in $\phi_{\rm sep}$, error propagation dictates that:
\begin{equation}
\label{eqn:error}
\frac{\Delta q}{q} = 5 \sqrt{0.002 + 0.001 \ln^2\left(\frac{\phi_{\rm sep}}{102\degree}\right) + \frac{\Delta \phi_{\rm sep}^2}{\phi_{\rm sep}^2}} \,.
\end{equation}

While this establishes a basis for measuring a planet's mass using the disk's spiral structure, we need to further investigate whether the secondary arm is observable, i.e., whether it is bright enough to be detected. For this purpose, the $\Sigma$ maps of \figref{fig:montage} can be misleading, because, as discussed by \citet{JBP2015}, variations in $\Sigma$ provides weak observability in comparison to variations in the vertical structure of the disk. Moreover, \citet{Zhu2015} has suggested that vertical linear perturbation modes are important at high altitudes, which can lead to enhancements in density perturbation by orders of magnitude compared to the midplane. These results point toward the importance of 3D effects, which we address in the following section.

\subsection{3D effects and observability of the secondary arm}
\label{sec:3D}

\begin{figure}[]
\centering
\includegraphics[width=0.99\columnwidth]{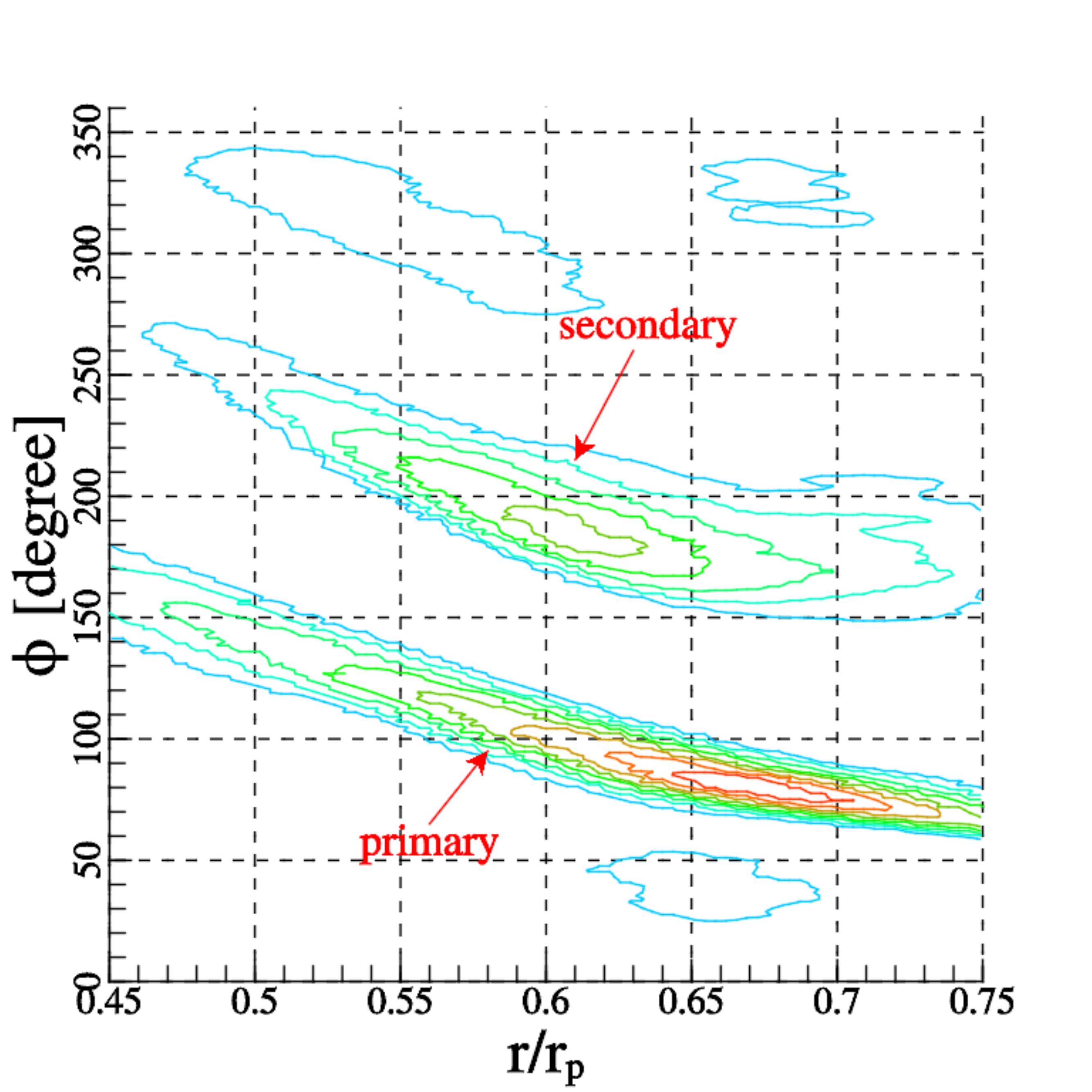}
\caption{Contour plot of the right panel in \figref{fig:scat_light}, in polar coordinates. The two arms are clearly seen in this plot, with the lower one being the primary arm. The separation between the two arms is nearly constant with an average of $96\degree$.}
\label{fig:cont}
\end{figure}

To investigate the differences between 2D and 3D disks, we additionally perform a 3D simulation for our $q=0.001$ and $H_0=0.1$ model. It is in spherical coordinates, with a polar angle $\theta$ extending from $0$ to $0.675$ radians above the midplane. We only simulate the upper half of the disk and assume midplane symmetry. The overall resolution is reduced by about $30\%$ to $288(r)\times 648(\phi)\times 72(\theta)$. In 3D, $r_{\rm s}$ is no longer required to be of order the disk scale height, and should be a small finite value only to avoid singularity. We therefore choose $r_{\rm s} = 0.1~r_{\rm H}$. The rest of the setup is identical to our 2D simulations.

After the simulation has reached 10 $P_{\rm p}$, we feed the simulation result to the \citet{Whitney2013} Monte-Carlo radiative transfer (MCRT) code to generate an $H$-band scattered light image with an angular resolution achievable by VLT/SPHERE, Gemini/GPI, and Subaru/SCExAO. The MCRT calculations largely follow the procedures presented in \citet{Dong2015b, Dong2015a}, and is only briefly repeated here. We scale the hydrodynamics model so that the planet is at 125 AU. The central source is a typical 1 $M_\odot$ star at 1 Myr with a temperature of 4350 K and a radius of 2.325 $R_\odot$. Dust grains are assumed to be ISM grains and well mixed with the gas (i.e., volume density of the dust is linearly proportional to the gas). The total mass of the ISM dust is assumed to be 2$\times10^{-5}~M_{\odot}$. We experimented with different choices of total dust mass between $10^{-5}$ to $10^{-4}~M_{\odot}$ and different filters including $Y$, $J$, $H$ and $K$-band. We find the measurement of $\phi_{\rm sep}$ does not depend on these choices.

Full resolution synthesized polarized intensity images at $H$-band are produced from the MCRT calculations, and they are convolved by a Gaussian point spread function with a full width half maximum (FWHM) of 0.05$\arcsec$ to achieve an angular resolution comparable with the current capability of Subaru, VLT, and Gemini. Other than the intrinsic Poisson noise of the Monte-Carlo method, we do not include any additional noise to the synthesized image. The target is assumed to be at 140 pc, a typical distance of nearby star formation regions. 

\figref{fig:scat_light} gives a side-by-side comparison for the surface density maps generated from the 2D and 3D hydrodynamics simulations (left and middle panels) along with the scattered light image translated from our 3D simulation (right panel). We find near perfect agreement between the surface density profiles of our 2D and 3D simulations. This is expected since at a large distance from the planet, the disk is effectively thin. In the scattered light image, however, some differences can be identified. First, the tertiary arm is too faint to be seen. Second, the pitch angles of both arms appear to be larger than their counterparts in surface density maps. This is due to the fact that scattered light comes from the disk surface, while surface density is determined mostly by the distribution of material at the disk midplane, and also the phenomenon observed by \citet{Zhu2015} --- a bending of the propagating shock wave that points toward the star at high altitudes. Since both arms are subjected to this effect, it does not noticeably affect $\phi_{\rm sep}$. Third, and most importantly, the secondary arm has a comparable brightness to the primary in the scattered light image, even though the surface density enhancement is much higher in the primary arm. This strongly suggests there is significant vertical motion near the arms, providing kinetic, rather than pressure, support to the local vertical structure, allowing the scattering surface of the secondary arm to rise to a similar height as the primary. A future study is required to thoroughly understand this feature.

One aspect of 3D effects that we have not taken into account in this study is the presence of a vertical temperature gradient. In a thermally stratified disk, density waves can become either deflected toward the disk surface, or confined in the midplane, depending on the direction of the temperature gradient \citep{Lubow1998,Lee2015}. Since we do not expect the primary and secondary arms to behave differently in this process of wave channeling, we do not expect $\phi_{\rm sep}$ to be dependent on this effect. However, we do expect it to have a significant impact on the brightness of the arms, making them more prominent if they are channeled toward the surface, or dimmer vice versa.

Finally, we measure $\phi_{\rm sep}$ in the scattered light image. As shown in \figref{fig:cont}, the two arms have a near constant azimuthal separation, and we find the average value of $\phi_{\rm sep}$ between $r/\rp = \{0.45,~0.65\}$ to be $\sim 96\degree$. Inserting this value into \eqnref{eqn:bestfit}, it predicts $q$ to be about $7.4\times 10^{-4}$, which successfully recovers the planet's mass to within $30\%$.

\subsection{Example: Application to SAO 206462}

Due to the simplicity of our scaling relation, given an image of the spirals, one can predict the mass of the planet responsible for the spiral structure with minimal effort. Here we use SAO 206462 as an example.

The circumstellar disk around SAO 206462 is a convenient choice to apply our results because it is nearly face-on with an inclination of only $11\degree$ \citep{Muto2012}. Therefore we may neglect projection effects and directly measure the separation of the two arms using Figure 1 of \citet{Garufi2013}. We find $\phi_{\rm sep} \sim 130\degree$. This translates to $q\sim0.0034$ using \eqnref{eqn:bestfit} to within $30\%$ accuracy, or $\sim6$ Jupiter masses, given the star's mass is $\sim1.7~M_\odot$ \citep{Muller2011}. We can also infer that the arm on the east side is the primary arm; therefore, this planet should be located toward the tip of that arm.

\section{Conclusions and Discussions}

We performed a parameter study for the azimuthal separation between the primary and secondary spiral arm, $\phi_{\rm sep}$, in the multi-armed spiral structure generated by a single planet on a fixed circular orbit, over a range of planet-to-star mass ratio, $q$, from $6.25\times 10^{-5}$ to $6.4\times 10^{-2}$, and disk aspect ratio at the planet's orbit, $H_0$, from $0.04$ to $0.16$. We found an empirical scaling that allows one to infer $q$ directly from $\phi_{\rm sep}$ (\eqnref{eqn:bestfit}) for planet mass companions; and also found that $\phi_{\rm sep}\sim 180\degree$ for brown dwarf mass companions. Using a near-infrared scattered light image synthesized by a combination of Monte-Carlo radiative transfer and 3D hydrodynamics simulation, we showed that this method is capable of measuring the mass of a Jupiter mass planet to within 30\% when the disk is face-on. Finally, we applied our scaling relation to SAO 206462, the most face-on two-armed spiral so far, and inferred that it contains a planet with about 6 Jupiter masses. In addition to observational applications, the spiral morphology described in this paper can serve as both motivation and pointers for further theoretical studies. In the following, we briefly discuss the implications of our results on nonlinear disk-planet interaction theory.

In the beginning of this paper we suggested that ultraharmonic density waves may be responsible for the multi-armed spiral structures. Even if this is true, without an in-depth theoretical understanding, it is difficult to predict what $\phi_{\rm sep}$ should be, since the multi-armed structure is the result of the interference between both the linear and ultraharmonic density waves, and to evaluate it requires precise knowledge of how the amplitudes of different azimuthal modes compare to each other. Nonetheless, it is surprising that $\phi_{\rm sep}$ depends on $q$ only, and not the disk temperature, because the linear density perturbation, which drives ultraharmonic waves, certainly depends on the disk temperature as well as $q$. Adding further complication to the picture, the location of the primary arm also changes with planet mass in the nonlinear regime, due to the spiral wave steepening into a shock \citep{Rafikov2002}. It may well be that to explain the scaling we found in this paper, both ultraharmonic waves and nonlinear wave propagation need to be taken into account.

\acknowledgments JF gratefully acknowledges support from the Center for Integrative Planetary Science at the University of California, Berkeley. We thank Pawel Artymowicz for computational resources and helpful feedback. We also thank Eugene Chiang for fruitful discussions during the process of writing this manuscript.

\clearpage\end{CJK*}
\end{document}